\documentclass[preprint]{aastex}
\usepackage[nolists,noheads,nomarkers]{endfloat}

\renewcommand\bf{}

\usepackage{longtable}
\usepackage{mathrsfs}

\usepackage{multirow}
\received{}
\revised{}
\accepted{}
\shorttitle{CO--0.30--0.07: A Peculiar Molecular Clump with an Extremely Broad Velocity Width}
\shortauthors{Tanaka et al.}
\tighten
\usepackage{float}

\newcommand\pcc{\mathrm{cm^{-3}}}

\newcommand\kelvin{\mathrm{ K}}

\newcommand\kmps{\ifmmode\mathrm{km\,s^{-1}}\else$\mathrm{km\,s^{-1}}$\fi}

\newcommand\ergs{\mathrm{ergs}}
\newcommand\pc{\mathrm{pc}}

\newcommand\Jypb{\ifmmode{\mathrm{Jy\,beam^{-1}}}\else{$\mathrm{Jy\,beam^{-1}}$}\fi}
\newcommand\mJypb{\ifmmode{\mathrm{mJy\,beam^{-1}}}\else{$\mathrm{mJy\,beam^{-1}}$}\fi}

\newcommand\Msol{\ifmmode{M_\odot}\else{$M_\odot$}\fi}

\newcommand\kB{{k_{\rm B}}}

\newcommand\Tkin{\ifmmode{T_{\rm kin}}\else{$T_{\mathrm{kin}}$}\fi}
\newcommand\Td{\ifmmode{T_{\mathrm{d}}\else{$T_{\mathrm{d}}$}\fi}}

\newcommand\Trot{\ifmmode{T_{\rm rot}}\else{$T_{\rm rot}$}\fi}

\newcommand\nHH{\ifmmode{n_{\rm H_2}}\else{$n_{\rm H_2}$}\fi}
\newcommand\ncrit{\ifmmode{n_{\rm crit}}\else{$n_{\rm crit}$}\fi}
\newcommand\ncritturb{\ifmmode{n_{\rm crit,turb}}\else{$n_{\rm crit,turb}$}\fi}
\newcommand\ncritshear{\relax\ifmmode{n_{\rm crit,tidal}}\else{$n_{\rm crit,tidal}$}\fi}

\newcommand\vlsr{\ifmmode{v_{\rm LSR}}\else${v_{\rm LSR}}$\fi}
\newcommand\Mvt{\ifmmode{M_{\rm VT}}\else${M_{\rm VT}}$\fi}

\newcommand\dv {{\rm d}v}
\newcommand\Eu {{E_{\rm u}}}
\newcommand\vexp{{v_{\rm exp}}}
\newcommand\vcol{\ifmmode{v_{\rm col}}\else{$v_{\rm col}$}\fi}

\newcommand\HCN{{\rm HCN}}
\newcommand\HCOp{\ifmmode{\rm HCO^+}\else{$\mathrm{HCO^+}$}\fi}
\newcommand\HCNt{\ifmmode{\rm H{^{13}C}N}\else{$\mathrm{H{^{13}C}N}$}\fi}

\newcommand\COt{\ifmmode{\rm {^{13}CO}}\else{$\mathrm{^{13}CO}$}\fi}
\newcommand\Ct{\ifmmode{\rm {^{13}C}}\else{$\mathrm{^{13}C}$}\fi}

\newcommand\NNHp{\ifmmode{\rm N_2H^+}\else{$\mathrm{N_2H^+}$}\fi}
\newcommand\HHCS{\ifmmode{\rm H_2CS}\else{$\mathrm{H_2CS}$}\fi}
\newcommand\CtS{\ifmmode{\rm ^{13}CS}\else{$\mathrm{^{13}CS}$}\fi}
\newcommand\OCS{\ifmmode{\rm OCS}\else{$\mathrm{OCS}$}\fi}
\newcommand\methanol{\ifmmode{\rm CH_3OH}\else{$\mathrm{CH_3OH}$}\fi}

\newcommand\JJ[2]{\ifmmode{\mbox{{\it J}={#1}\mbox{--}{#2}}}\else{{\it J}={#1}--{#2}}\fi}
\newcommand\JK[4]{\ifmmode{{J_K}=#1_{#2}\mbox{--}#3_{#4}}\else{${\it J_K}=#1_{#2}\mbox{--}#3_{#4}$}\fi}

\newcommand\CIa{\ifmmode{^3}P_1\mbox{--}{^3}P_0\else${^3}P_1\mbox{--}{^3}P_0$\fi}

\newcommand\gl{l}
\newcommand\gb{b}

\newcommand\CLb{CO$-0.30$$-0.07$}
\newcommand\theObj\CLb

\newcommand\Dv{\ifmmode{\Delta v}\else{$\Delta v$}\fi}
\newcommand\vc{\ifmmode{\left<v\right>}\else{$\left<v\right>$}\fi}

\newcommand{\PV}{{\it P}--{\it V}}
\newcommand{\avir}{\ifmmode{\alpha_{\rm vir}}\else{$\alpha_{\rm vir}$}\fi}

\newcommand{\meta} {\ifmmode{{6_{1}\mbox{--}{5_{2}}}\,\mathrm{E}}\else{${6_{1}}$--${5_{2}}$\,E}\fi}
\newcommand{\metb} {\ifmmode{{5_{2}\mbox{--}{4_{1}}}\,\mathrm{E}}\else{${5_{2}}$--${4_{1}}$\,E}\fi}
\newcommand{\metc} {\ifmmode{{9_{0}\mbox{--}{8_{1}}}\,\mathrm{E}}\else{${9_{0}}$--${8_{1}}$\,E}\fi}
\newcommand{\metd} {\ifmmode{{9_{-1}\mbox{--}{8_{0}}}\,\mathrm{E}}\else{${9_{-1}}$--${8_{0}}$\,E}\fi}

\begin{document}

 
\title{CO--0.30--0.07: A Peculiar Molecular Clump with an Extremely Broad Velocity Width in the Central Molecular Zone of the Milky Way}
\author{Kunihiko Tanaka}
\email{ktanaka@phys.keio.ac.jp}
\affil{Department of Physics, Faculty of Science and Technology, Keio University, 3-14-1 Hiyoshi, Yokohama, Kanagawa 223--8522 Japan}

\author{Makoto Nagai}
\affil{Division of Physics, Faculty of Pure and Applied Sciences,  University of Tsukuba, Ten-noudai 1-1-1, Tsukuba, Ibaraki 305--8571 Japan}

\author{Kazuhisa Kamegai}
\affil{Department of Industrial Administration, Faculty of Science and Technology, Tokyo University of Science, 2641 Yamazaki, Noda, Chiba 278--8510 Japan}

\and 

\author{Tomoharu Oka}
\affil{Department of Physics, Faculty of Science and Technology, Keio University, 3-14-1 Hiyoshi, Yokohama, Kanagawa 223--8522 Japan}



\begin{abstract}
The high velocity {\bf dispersion} compact cloud CO$-0.30$$-0.07$ is a peculiar molecular clump discovered in the central moleculr zone of the Milky Way, which is characterized by its extremely broad  velocity emissions ($\sim 145\ \rm{km\,s^{-1}}$) despite the absence of internal energy sources.
We present new interferometric maps of the cloud in multiple molecular lines in frequency ranges of 265--269 GHz and 276--280 GHz obtained using the Sumbmillimeter Array, along with the single-dish images previously obtained with the ASTE 10-m telescope.
The data show that the characteristic broad velocity emissions are predominantly confined in two parallel ridges running through the cloud center.
The central ridges are tightly anti-correlated with each other in both space and velocity, thereby sharply dividing the entire cloud into two distinct velocity components (+15 km\,s$^{-1}$ and +55\ km\,s$^{-1}$).
This morphology is consistent with a model in which the two velocity components collide with a relative velocity of 40 \kmps\ at the interface defined by the central ridges, although an alternative explanation with a highly inclined expanding-ring model is yet to be fully invalidated. 
We have also unexpectedly detected several compact clumps ($\lesssim 0.1\ $pc in radius) likely formed by shock compression.
The clumps have several features in common with typical star-forming clouds: high densities ($10^{6.5\mbox{--}7.5}\ \mathrm{cm^{-3}}$), rich abundances of hot-core-type molecular species, and relatively narrow velocity widths apparently decoupled from the furious turbulence dominating the cloud.
The cloud CO$-0.30$$-0.07$ is possibly at an early phase of star formation activity triggered by the shock impact.

\end{abstract}

%
%

\section{INTRODUCTION}
Widespread strong turbulence is an outstanding characteristic of the molecular clouds (MCs) in the central molecular zone (CMZ) of the Milky Way.
Their velocity dispersions lie about a factor of 5 above the size--velocity width relationship for the Galactic disk, irrespective of choice of tracer lines \citep{Miyazaki2000,Oka2001a,Shetty2012a}.
The virial parameters ($\avir$) of the CMZ clouds are systematically $\sim 10$ times those in the Galactic disk, indicating that they are not self-gravitating but likely bound by external pressure \citep{Miyazaki2000,Oka2001a}. 
{\bf
It is suggested by many authors that the high degree of turbulence in the CMZ may affect the star formation process there. 
Primarily, the strong turbulence suppresses the star formation rate (SFR) because the gravitational collapse of dense clumps in the MCs is slowed or inhibited by the high turbulent pressure.
Several recent works suggest that the SFR within the CMZ may be suppressed by this effect to a value 1--2 orders of magnitude lower than the value expected from its dense gas mass \citep{Longmore2012b,Kruijssen2014}, although other effects such as the tidal shear, strong magnetic field, and enhanced cosmic-ray flux may also play significant roles in determining the SFR \citep{Yusef-Zadeh2007,Kruijssen2014}.
Meanwhile, turbulence may also facilitate the formation of massive stars and stellar clusters in sites of cloud--cloud collision \citep{Habe1992,Inoue2013a}.  
This collision-induced star formation is a popular explanation for the highly active star formation in the Sgr B2 complex \citep{Hasegawa1994}, and there is increasing evidence suggesting that cloud--cloud collisions play a key role in cluster formation in the CMZ \citep{Higuchi2014} as well as in the Galactic disk region \citep{Furukawa2009,Higuchi2010a}.
}

The properties and origin of turbulence in the CMZ are, however, yet to be fully understood.
{\bf
\cite{Shetty2012a} show that the size--linewidth relationship holds both within and among the CMZ clouds over an order of magnitude in the spatial scale.
This indicates that the turbulence driving in the CMZ is predominantly a large scale one, which is consistent with the idea that the turbulence is maintained by interactions with frequent supernova explosions \citep{Shetty2012a} or with the bar potential in the inner Galaxy \citep{Rodriguez-Fernandez2006}.}
Meanwhile, high spatial resolution surveys in millimeter and submillimeter carbon monoxide (CO) lines \citep{Oka1999,Oka2012}\ detected a considerable number of small clumps with extremely broad velocity widths inconsistent with the size--velocity width relationship; 
their velocity widths are 40--120 $\kmps$ in full width at zero-intensity (FWZI) in spite of their small spatial sizes of approximately 1--3 pc in radius, indicating enhancement in velocity widths by a factor of 4  from the size--linewidth relationship \citep{Tanaka2014}.
{\bf
These clumps are distinguished by their spatial compactness from the larger-scale broad-line structures of a few 10 pc in sizes that were identified by \cite{Bania1977} and \cite{Liszt2006} at larger Galactic longitudes ($|l|>1^\circ$).
}
The recent CO \JJ{3}{2}\ survey with the ASTE 10-m telescope \citep{Oka2012}\ identified 70 such high velocity {\bf dispersion} compact clouds \citep[HVCCs;][]{Oka2007} and HVCC-like features, most of which were not associated with known energy sources for driving such small scale turbulence.

Recently, we reported results of the HCN \JJ{4}{3}\ mappings obtained with the ASTE 10-m telescope toward the Sgr C complex, which include one of the most prominent HVCCs in the CMZ, \theObj\ \citep{Tanaka2014}.
{\bf We showed that the cloud had a velocity width of 120 \kmps\ comprising a pair of broad velocity lobes with a velocity width of 60 \kmps\ each, which were anti-correlated with each other in both space and velocity. }
The kinetic energy of the entire system was estimated to be $10^{49}$ ergs, which is comparable to those of the clouds interacting with supernova remnants (SNRs).
We proposed several hypotheses for the origin of the HVCC: cloud--cloud collision, hyper-energetic molecular outflow, a compact expanding shell, and a rotating ring/disk, although the spatial resolution of the telescope ($24''$ = 1 pc) was insufficient to distinguish these possible cases.

{\bf
This paper reports the results of new observations toward \theObj\ performed with the Submillimeter Array (SMA).
We present interferometric images of the cloud in submillimeter lines of HCN, \HCOp, \NNHp, \methanol, and several other hot-core-type molecules probing the dense molecular gas ($n \gtrsim 10^5\ \pcc$) in a wide range of physical and chemical conditions.
The HCN, \HCOp, and \NNHp\ lines are commonly used high density tracers, which are sensitive to different environments; \NNHp\ tends to selectively trace quiescent, cold cores \citep[e.g.][]{Tatematsu2008}, whereas HCN and \HCOp\ abundances are less environmentally dependent and hence they are often useful in detecting shock-compressed gas \citep[e.g.][]{Tanaka2007,Tanaka2014}. 
Methanol is one of the most abundant interstellar organic molecules with enhanced abundance in star-forming cores, though other enhancement mechanisms such as cosmic-ray-induced photo-desorption \citep{Yusef-Zadeh2013b} and ejection by shock \citep{Requena-Torres2006} are also suggested to be important in the CMZ.
}
With these data, we investigate the structure of the shocked gas in the cloud in detail and thereby examine whether it shows systematic motion such as collimated outflow, expansion, and rotation, or its large velocity dispersion is caused by random turbulent motion.
We also report the unexpected discovery of compact clumps toward the boundary between the two velocity components of the cloud, which may indicate an early phase of star formation activity under the highly turbulent environment.
In this paper, we adopt 8.3 kpc for the distance to the Galactic center \citep{Gillessen2009}.

\section{OBSERVATIONS}

\subsection{SMA Observations}
The observations were performed on May 12th, 2013, by using the Submillimeter Array (SMA) in the sub-compact configuration providing baseline lengths ranging from 9.5 m to 25 m.
We conducted a hexagonal 7-point Nyquist-sampled mosaic to cover the $52''$ radius region centered at $\left(\alpha_{2000}, \delta_{2000}\right) = (\mathrm{17^h45^m08^s.0}, \mathrm{-29^\circ13'44''.8})$ or $\left(\gl, \gb\right) = \left(-0^\circ.3053, -0^\circ.0622\right)$ in the Galactic coordinates.
The 230 GHz receivers were tuned to observe frequency ranges from 264.77 GHz to 268.75 GHz (LSB) and from 276.77 GHz to 280.75 GHz (USB) simultaneously. 
The correlator configuration was chosen to provide  4 GHz frequency coverage in each sideband and 812.5 kHz spectral resolution.
The HCN \JJ{3}{2}, \HCOp\ \JJ{3}{2}, and \methanol\ $J_K$=\meta, \metb, and \metc\ lines were observed in the LSB, and the \NNHp\ \JJ{3}{2}, \CtS\ \JJ{6}{5}, \OCS\ \JJ{23}{22}, o-\HHCS\ $J_{K_{\rm a},K_{\rm c}}=8_{1,7}$--$7_{1,6}$, and \methanol\ $J_K$=\metd\ lines were in the USB.  
We list the target lines in Table \ref{TABLE1}, along with their rest frequencies and  upper state energies.

We observed Neptune, 3C279, and nrao530 for the flux, bandpass, and gain calibration measurements, respectively.
We employed the CASA package developed for the NRAO for flagging bad data, applying calibrations, and creating images.
We applied Briggs weighting for the imaging, which provided a $8''.2\times5''.2$ synthesized beam for HCN \JJ{3}{2}.
The beam position angle is $-0^\circ.5$ in the equatorial coordinates or $-59^\circ.1$ in the Galactic coordinates.
Phase-only self-calibration was applied to the quasi-continuum data in each sideband, the calibration table of which was transferred to the line data.
The continuum image was constructed by integrating the emissions of the line-free channels in the total 8 GHz bandwidth.
The total on-source integration time was 25 min per pointing, and the resultant r.m.s. noise level of the final maps is 0.22 \Jypb\ per velocity channel for the USB, 0.27 \Jypb\ for the LSB, and 3.9 \mJypb\ for the continuum data. 
Correction for the primary beam was not applied.  
The maps were originally made in the equatorial coordinate system, and subsequently we converted them into the Galactic coordinate system for convenience of comparison with the single-dish HCN \JJ{4}{3}\ data obtained with the ASTE 10-m telescope \citep{Tanaka2014}.

\begin{deluxetable}{lcccc}
\tablecaption{Molecular Lines Observed with the {\bf SMA}}


\tablecolumns{5}
\small
\tablehead{ \colhead{Molecule} & \colhead{Transition} & \colhead{Frequency} & \colhead{$\Eu$/$\kB$} & \colhead{sideband}\\ \colhead{} & \colhead{} & \colhead{(GHz)} & \colhead{(K)} & \colhead{USB/LSB} }
\tabletypesize{\scriptsize}
\tablewidth{0pt}
\startdata
\HCN     & 3--2 & 265.886 & 25.5 & L \\
\HCOp     & 3--2 & 267.558 & 25.7  & L\\
\NNHp     & 3--2 & 279.512 & 26.8 & U \\
\CtS      & 6--5 & 277.455 & 23.1 & U \\
o-\HHCS & $8_{1,7}$--$7_{1,6}$ & 278.888 &  73.4 & U \\
\OCS    & 23--22 & 279.685 & 161.1 & U \\
\methanol & ${6_{1}}$--${5_{2}}$ E & 265.290 & 61.9 & L  \\
          & ${5_{2}}$--${4_{1}}$ E & 266.838 & 49.2 &L \\
          & ${9_{0}}$--${8_{1}}$ E & 267.403 & 109.6 &L\\
          & ${9_{-1}}$--${8_{0}}$ E & 278.305 & 102.1 &U 
\enddata

\label{TABLE1}
\end{deluxetable}

\subsection{ASTE Observations}
We also use the published data of HCN $\JJ{4}{3}$ and $\COt\ \JJ{3}{2}$ obtained with the ASTE 10-m telescope \citep{Tanaka2014}.
The maps were obtained with $24''$ spatial resolution and $2\ \kmps$ velocity resolution by employing the On-The-Fly (OTF) scan mode.
The r.m.s. noise levels in the main-beam temperature scale are 0.13 K and 0.21 K for the HCN and \COt\ data, respectively.
Further details of the ASTE data and observations are described in \cite{Tanaka2014}.


\section{RESULTS}
\begin{figure*}[hhh]
\epsscale{1}
\plotone{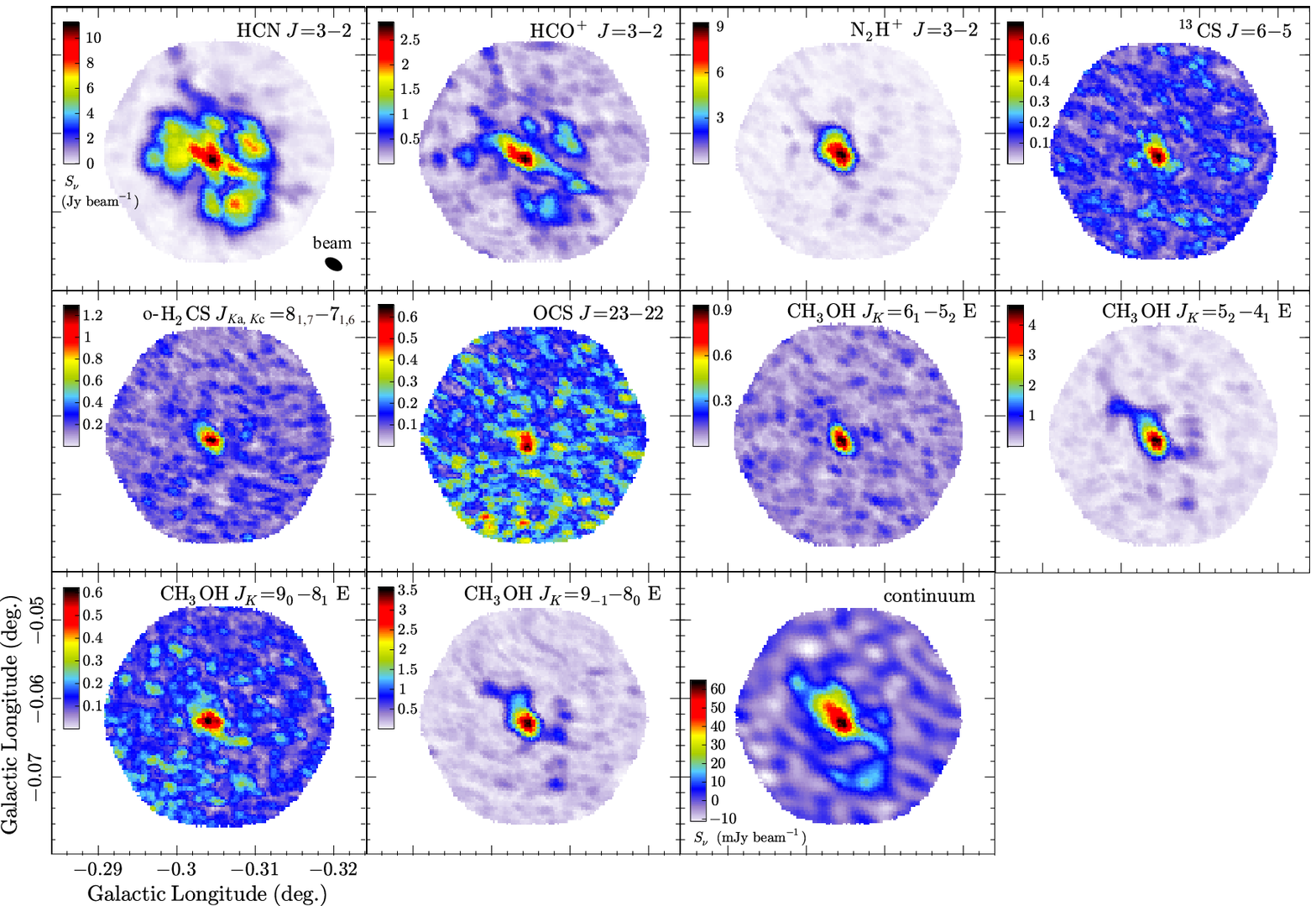}
\caption{Maps of the peak flux densities of the observed lines and the continuum flux densities.  The peak flux densities are calculated from spectra binned over 10 velocity channels. The synthesized beam is represented by a filled ellipse on the HCN \JJ{3}{2}\ panel.  } \label{FIG1}
\end{figure*}

\begin{figure*}[hhh]
\epsscale{1}
\plotone{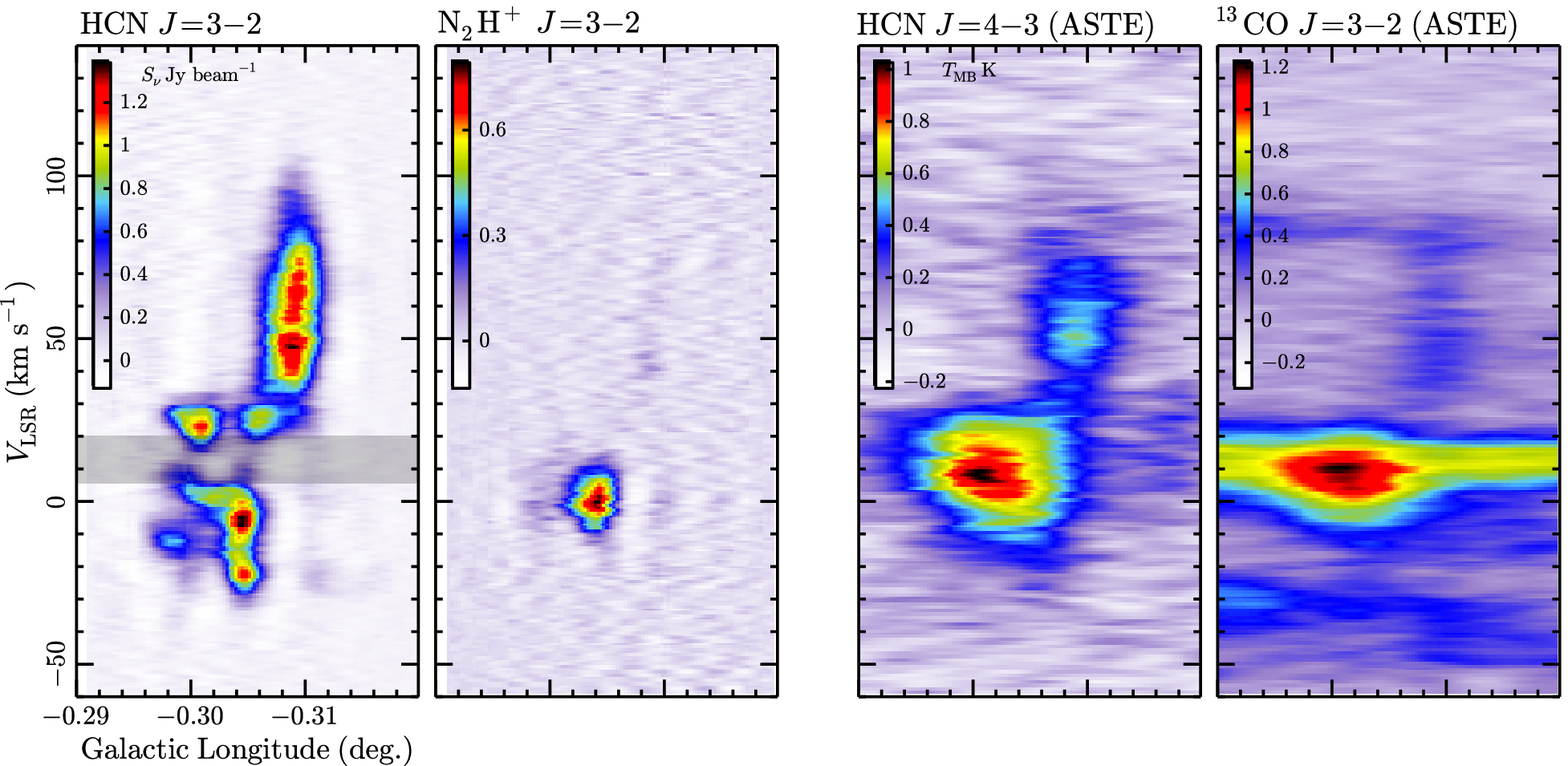}
\caption{Galactic longitude--velocity diagram of \theObj\ averaged over the full latitude range in HCN \JJ{3}{2}\ and \NNHp\ \JJ{3}{2}\ obtained with the SMA observations, along with the HCN \JJ{4}{3}\  and \COt\ \JJ{3}{2}\ images obtained with the ASTE observations.  The dark hatched region on the leftmost panel indicates the velocity range affected by the artificial velocity gap in the synthesis observation.} \label{FIG2}
\end{figure*}

\subsection{Overall Structure: Central Clump and Extended Turbulent Component}
Figure \ref{FIG1} shows the peak flux densities of the observed lines and continuum flux density, where the peak flux densities were calculated from spectra binned over 10 velocity channels to improve S/N ratios.
Two spatial components can be recognized in the maps. 
One is the bright compact clump toward the field center, which is commonly seen in all the observed lines and  the continuum but more clearly in the maps of the optically thinner group (i.e., the continuum and the lines other than HCN and \HCOp).
This central clump is point-source-like in the methanol, \OCS, \CtS, and o-\HHCS\ maps, whereas its \NNHp\ and continuum images have sizes significantly larger than the beam size.  
The other component is the clumpy, spatially extended component within a radius of approximately $30''$ around the central clump.
This extended component is dominant in the HCN and \HCOp\ maps and is recognizable in the maps of \NNHp\ and \methanol\ \metb\ and \metd\ as well as the continuum.  
The mass distribution is better represented by the optically thin group, indicating that the major fraction of the cloud mass is confined in the central clump.

The contrast between the central clump and the extended component is more striking in Figure \ref{FIG2}, which shows the position--velocity (\PV) diagrams along the Galactic latitude axis for the SMA data of the HCN and \NNHp\ lines averaged over the full latitude range.
The extended component in the HCN map shows a highly turbulent velocity structure with emission spanning in a \vlsr\ range of 145 \kmps\ from $-35\ \kmps$ to $+110\ \kmps$, which is wider than that previously measured with our single-dish observations.
In contrast, the central clump in the \NNHp\ image is strictly limited to a narrow velocity range from $-10\ \kmps$ to $+10\ \kmps$, apparently decoupled from the outer highly turbulent flow. 
The HCN image has a peak toward the central clump at a velocity slightly blue-shifted by $5\ \kmps$ presumably because of self-absorption.   
We also show the single-dish HCN \JJ{4}{3}\  and \COt\ \JJ{3}{2}\ data in Figure \ref{FIG2}, in which only the extended component is visible; they do not have a narrow peak at the velocity corresponding to the central clump.

The emission gap of 15 \kmps\ width at $\vlsr\sim+15\ \kmps$ seen in the SMA HCN \JJ{3}{2}\ data in Figure \ref{FIG2} is likely to be an artifact {\bf resulting from the limited $uv$ coverage in our observations.}
The single-dish \COt\ map shows that the emission in this velocity range extends beyond the primary beam of the SMA, implying that it is resolved out in the synthesis observation.

\subsection{Double-lobed Structure: Tight Anti-correlation between the Broad Velocity Lobes}

\begin{figure*}[ttt]
\epsscale{1.}
\plotone{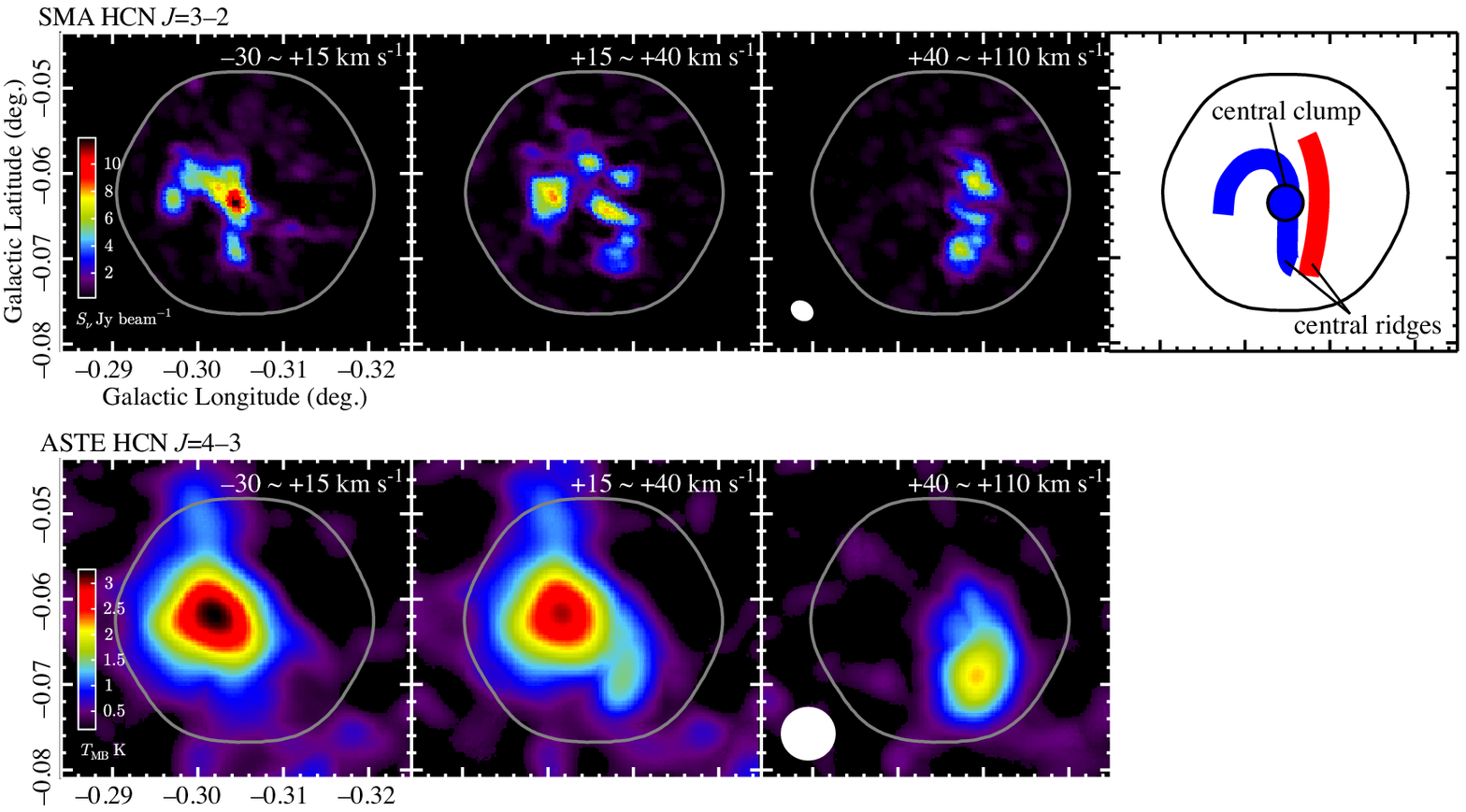}
\caption{Peak flux density/brightness temperature maps for the blue-shifted lobe (leftmost; $\vlsr$ range from $-35\ \kmps$ to $+15 \kmps$), intermediate velocity range (second from  left; from $+15$ \kmps to +40 \kmps), and red-shifted lobe (third from left; from +40 \kmps\ to +110 \kmps).  The top and bottom rows are for the SMA HCN \JJ{3}{2}\ and  ASTE \JJ{4}{3}\ data, respectively.  The rightmost panel of the upper row is a schematic diagram showing the positions of the central ridges in the blue- and red-shifted lobes (shown in blue and red, respectively) and the central clump.  The coverage of the SMA observation is superposed on the HCN \JJ{4}{3}\ maps.  }\label{FIG3}
\end{figure*}

\begin{figure*}[ttt]
\epsscale{1.}
\plotone{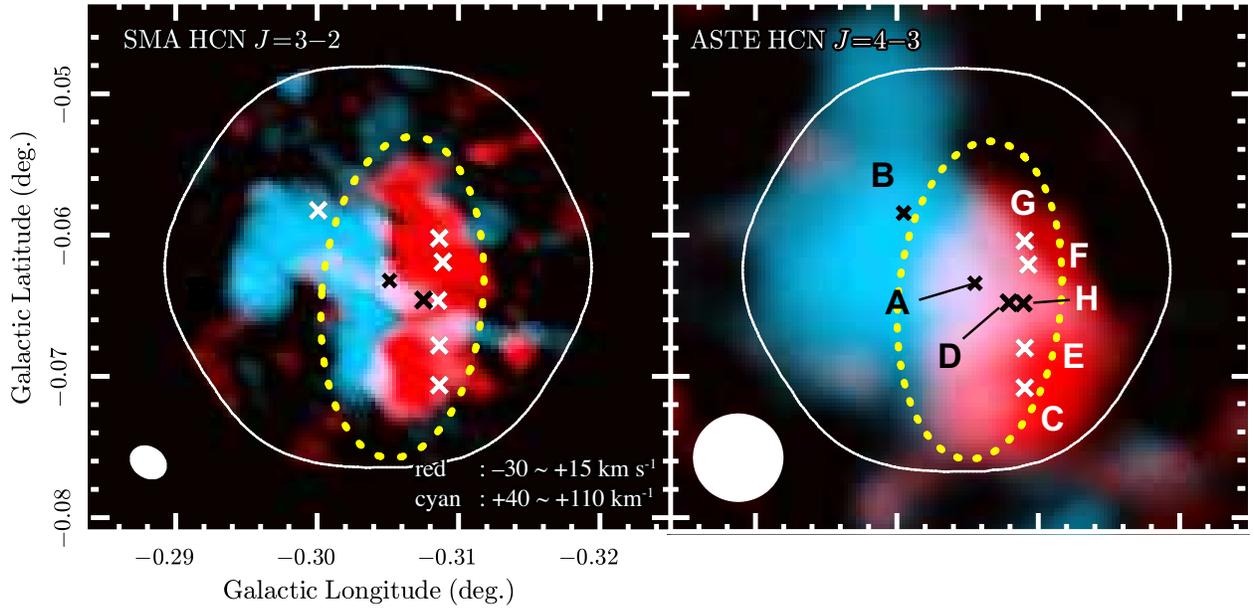}
\caption{Composite color images of the two broad velocity lobes, in which the $3\ \sigma$ emission regions of the blue-shifted and red-shifted lobes are shown in cyan and red, respectively, and overlapping regions are shown in white. The region corresponding to the central ridges is encircled by yellow dashed lines.  The methanol peaks A--H are indicated by cross marks.  }\label{FIG4}
\end{figure*}

\begin{figure}[ttt]
\epsscale{0.7}
\plotone{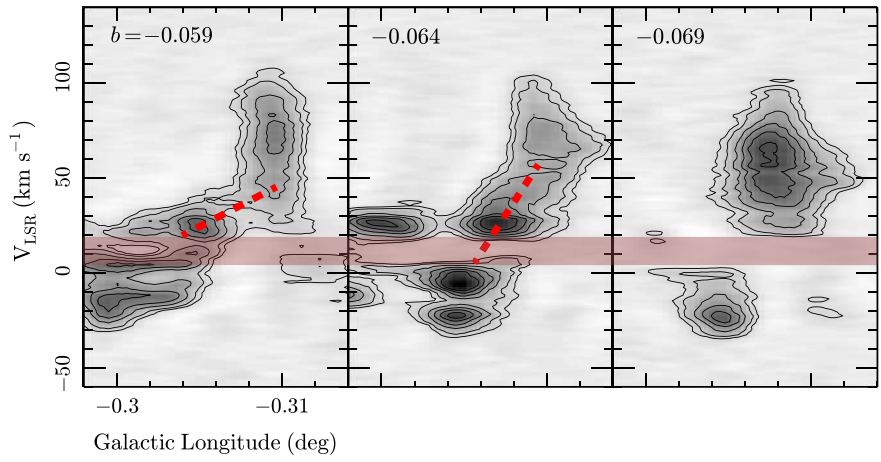}
\caption{HCN \JJ{3}{2}\ position--velocity diagrams along the horizontal strips passing thorough northern end ($b = -0^\circ.059$),  midpoint ($-0^\circ.064$), and southern end ($-0^\circ.069$) of the central ridges.  The bridging emission between the two broad velocity lobes are indicated by dashed lines. The artificial velocity gap in the synthesis observation is hatched in gray.}\label{FIG5}
\end{figure}

\begin{figure*}[ttt]
\epsscale{1.}
\plotone{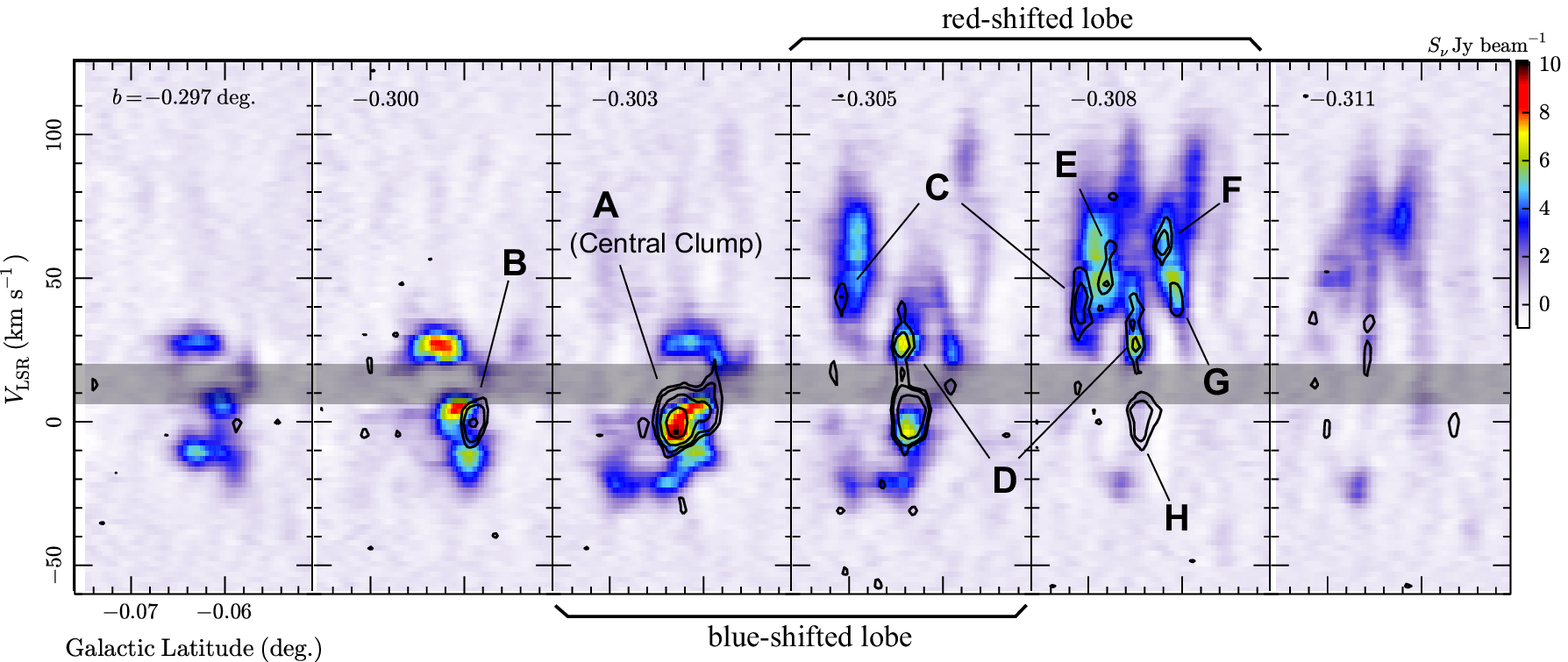}
\caption{Galactic latitude--\vlsr\ diagrams of HCN $\JJ{3}{2}$ at every $10''$ Galactic longitude, with contours of \methanol\ $J_K=\metd$ overlaid. The contour levels are $(3, 5, 10,$ and $ 20)\times \sigma$ ($\sigma = 0.10$ Jy\,beam$^{-1}$). The HCN and \methanol\ spectra are smoothed with Gaussian kernels of 2 \kmps and 5 \kmps\ FWHM, respectively.  Bright methanol peaks are labeled as A--H.  The hatched regions are for the artificial velocity gap in the synthesis observation. }\label{FIG6}
\end{figure*}

The double-lobed structure detected with the ASTE 10-m telescope observation is also clearly recognizable in the SMA image with the improved resolution. 
The \PV\ diagram of HCN \JJ{3}{2}\ presented in Figure \ref{FIG2} shows that the two broad velocity lobes are sharply separated in space at $\gl = -0^\circ.306$, with a narrow overlapping \vlsr\ range from +15\ \kmps\ to +35\ \kmps.
Figure \ref{FIG3} shows the peak flux density maps of the HCN \JJ{3}{2}\ for the $\vlsr$ ranges of the blue-shifted lobe (from $-35$ \kmps\ to $+15$ \kmps), the intermediate velocity range (from $+15\ \kmps$ to $+40\ \kmps$), and the red-shifted lobe (from $+40$\ \kmps\ to $+110$\ \kmps), along with the single-dish HCN $\JJ{4}{3}$ maps for the same velocity ranges.
The SMA images reveal that the two broad velocity lobes have a ridge-like shape; each broad velocity lobe has a vertical structure near the field center, as schematically shown in the upper rightmost panel.
Hereafter we refer to this pair of parallel ridges as the central ridges.
The emissions from the red-shifted lobe primarily arise from the central ridge part.
The blue-shifted lobe also shows bright emissions in the eastern region behind the central ridge, making the entire lobe appear inverse-`c' shaped.
The central clump is located toward the mid-point of the blue-shifted central ridge.
The emissions from the intermediate velocity range ($-15\ \kmps$ to $+40\ \kmps$) are mostly extensions from the two broad velocity lobes, and no notable spatial structures intrinsic to this velocity range are detected.

Figure \ref{FIG4} shows composite color images constructed from the two broad velocity lobes, where the 3 $\sigma$ emission regions of the blue- and red-shifted lobes are shown in cyan and red, respectively. 
As reported in \cite{Tanaka2014}, the two velocity lobes in the single-dish image show a good spatial anti-correlation, with a narrow overlapping region of an about half the beam size.
Now we recognize that the anti-correlation is much tighter in the interferometric images; 
the central ridges correspond exactly to the overlapping region in the single-dish image, but they have almost no spatial gaps and only few overlaps between each other. 
Thus the two broad velocity lobes appear to contact each other at a well-defined interface on the plane of the sky.
A natural explanation for this remarkable structure is that the two components with different velocities physically interact with each other, that is, they collide.
Meanwhile, our results do not completely exclude other interpretations, as the paired ridges also seem to compose one elongated ring. 
We will discuss this issue further in later sections (\ref{DISCUSSION1}, \ref{DISCUSSION2}).

The emissions of the central ridges partly mix in the intermediate velocity range.
Figure \ref{FIG5} shows the \PV\ diagrams of the SMA HCN \JJ{3}{2}\ data along horizontal paths across the northern end ($\gb = -0^\circ.59$),  midpoint ($\gb = -0^\circ.64$), and  southern end ($\gb = -0^\circ.69$) of the ridges.
We find several features bridging the velocity gap between the central ridges along these paths.
Along the central strip, the two velocity components are smoothly connected by emission with a velocity gradient in a \vlsr\ range from $-30$\ \kmps\ to +60\ \kmps, except for the artificial emission gap noted in the previous subsection.
A bridging feature is also detected along the northern strip in a narrower \vlsr\ range from +20\ \kmps\ to +40\ \kmps.
The southern strip lacks detectable emission in the intermediate velocity range, although it is possibly because of the missing flux.

\subsection{Broad HCN Emission}
Figure \ref{FIG6} shows the HCN \JJ{3}{2}\ \PV\ diagrams along the Galactic latitude at every $10''$ Galactic longitude, with contours of \methanol\ \metd\ emission overlaid, where the HCN and \methanol\ spectra are smoothed with Gaussian kernels of 2\ \kmps\ and 5\ \kmps\ FWHM, respectively.
The broad velocity lobes are resolved into a number of smaller broad emissions.
The red-shifted lobe ($\vlsr \gtrsim +40\ \kmps$) is found to comprise at least 6 broad features with about $10''$ sizes and velocity widths of 40 -- 50 \kmps\ in FWZI, the systemic velocities of which are distributed in a $\vlsr$ range from $+35\ \kmps$ to $+90\ \kmps$ without apparent order.   
The blue-shifted lobe ($\vlsr \lesssim +15\ \kmps$) does not seem to have prominent broad features but rather seems to be an ensemble of relatively narrow clumps, although this is at least partly because of the artificial emission gap at $\vlsr \sim +15\ \kmps$; any broad features in this velocity range would be unrecognizable as they would be divided by this artificial gap.
The narrow emissions in the $\vlsr$ range from $-30\ \kmps$ to $-20\ \kmps$ could be the low-velocity ends of such obscured broad features.

The broad emission features are predominantly associated with the central ridges mentioned in the previous subsection, which correspond to the Galactic longitude range from $-0^\circ.303$ to $-0^\circ.308$ in Figure \ref{FIG6}.
They are rather randomly distributed in the position-velocity space, in which we cannot detect definitive signs of systematic motion over the entire clump, such as coherent expansion, rotation, and linear collimated flow.  

We label 8 bright methanol peaks as A--H in Figure \ref{FIG6}, the positions of which on the $\gl$-$\gb$ plane are also shown in Figure \ref{FIG4}.
Peak A is the methanol peak in the central clump. 
All of the methanol peaks have velocity widths of $10$--$20\ \kmps$, which are narrower than those of the HCN emissions.
They are primarily associated with the broad HCN features in the central ridges, except for peak B, which is located well outside the ridges, and peak H, which is located inside the red-shifted ridge but has an inconsistent blue-shifted velocity.
{\bf
We list the positions of the methanol peaks in Table \ref{TABLE2}, along with their center LSR velocities and FWHM velocity widths determined by Gaussian fitting of the \metb\ spectra.
}

\begin{deluxetable}{lrrrrl}
\tablecaption{Methanol Peaks}



\tablecolumns{6}
\small
\tablehead{ \colhead{} & \colhead{{$l$}} & \colhead{{$b$}} & \colhead{{$\vlsr$}} & \colhead{{$\Delta{v}$}} & \colhead{Reference} \\
\colhead{} & \colhead{($^\circ$)} & \colhead{($^\circ$)} & {($\kmps$)} & {($\kmps$)} }
\tabletypesize{\scriptsize}
\tablewidth{0pt}
\startdata
A & $-0.305$ & $-0.064$ & $1.1\pm0.2$  & $12.8\pm0.3$ & Central Clump, 36 GHz maser candidate \tablenotemark{1} \\
B & $-0.300$ & $-0.058$ & $0.1\pm0.4$ & $12.1\pm0.9$ \\
C & $-0.309$ & $-0.071$ & $45.4\pm1.0$ & $22.0\pm2.4$ \\ 
D & $-0.308$ & $-0.065$ & $26.0\pm1.0$ & $23.7\pm2.5$ \\
E & $-0.309$ & $-0.068$ & $54.0\pm1.7$ & $17.7\pm4.5$ \\
F & $-0.309$ & $-0.062$ & $59.5\pm0.5$ & $6.8\pm1.2$  \\
G & $-0.309$ & $-0.060$ & $45.4\pm0.7$ & $13.6\pm1.8$ \\
H & $-0.309$ & $-0.065$ & $-0.2\pm0.9$ & $17.8\pm2.3$ 
\enddata
\tablenotetext{1}{\cite{Yusef-Zadeh2013b}}

\label{TABLE2}
\end{deluxetable}

\subsection{Central Clump}
\begin{figure}[ttt]
\epsscale{0.6}
\plotone{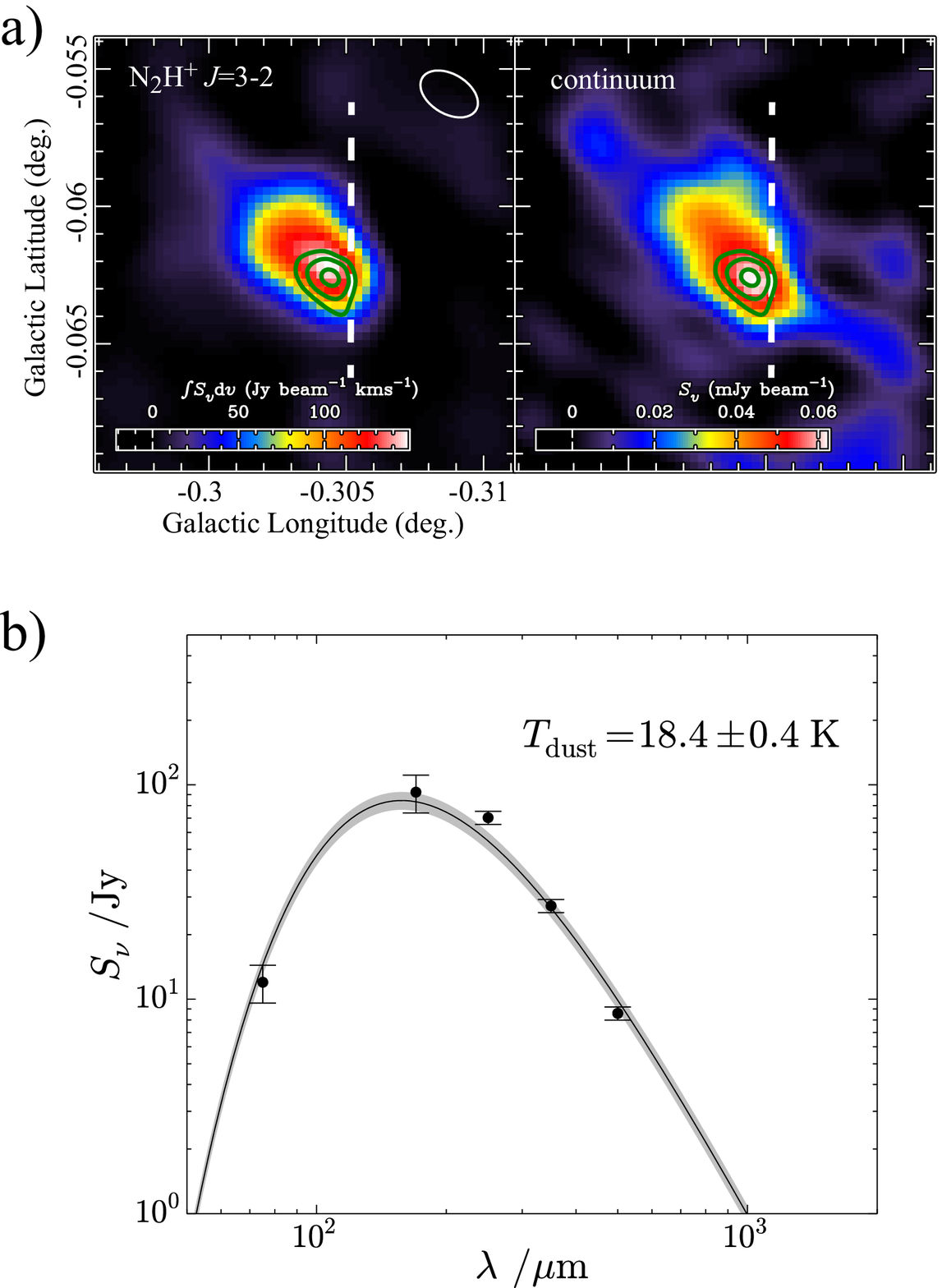}
\caption{(a) SMA images of the central clump in the continuum flux density (left) and the $\NNHp\ \JJ{3}{2}$ velocity-integrated flux density (right).  The integration range is from $-10\ \kmps$ to $+20\ \kmps$. The overlaid contours are the \methanol\ $J_K=\metc$ flux integrated over the same velocity range, drawn at every 1 $\sigma$ from 5 $\sigma$ ($\sigma = 1.5\ {\mathrm{Jy\,beam^{-1}}}\ \kmps$). The boundary between the central ridges is schematically drawn with a dashed line on each panel. (b) Spectral energy distribution of the central clump in the far-infrared wavelengths, constructed from 70 -- 500 $\micron$ flux density from the {\it Herschel} data archive. The best-fit gray-body curve (solid line) and 1-$\sigma$ deviation (hatched region) are overlaid. }\label{FIG7}
\end{figure}

Figure \ref{FIG7}a shows the SMA images of the central clump in the continuum, and the $\NNHp\ \JJ{3}{2}$\ and \methanol\ \metc\ flux densities integrated over a velocity range from $-10\ \kmps$ to $+20\ \kmps$.
{\bf Table \ref{TABLE3} summarizes the properties of the central clump.}
The clump has approximately the same elliptic shape in the continuum and in the \NNHp\ maps; they are elongated in the northeast--southwest direction with FWHM sizes of $13''\times8''$.
The point source-like methanol \metc\ emission is offset from the center position of the clump toward the south-western edge, i.e., toward the boundary between the two velocity components.

We estimate the dust luminosity mass of the central clump by referring to the {\it Herschel} archival data. 
First, we determine the dust temperature ($\Td$) by analyzing the spectral energy distribution (SED) constructed from the 70--500\ \micron\ flux densities taken from the {\it Herschel} data archive.
Figure \ref{FIG7}b shows the SED for the central $40''$ square region, where contributions from extended emission unassociated with the cloud are subtracted by taking the background position at $\pm60''$ offset from the cloud center in the Galactic longitude and latitude.
The error bars include the calibration errors in the {\it Herschel} data and uncertainties of the background levels estimated using the standard deviations among the flux densities toward the four reference positions.
The SED is consistently fitted by a single-component gray-body model with a dust temperature ($\Td$) of 18 K and the dust emissivity index ($\beta$) assumed to be 2.0.  
Although a simultaneous SED fitting of $\Td$ and $\beta$ suggests a higher $\beta$ value of 2.8, we fix $\beta$ at the conservative value of 2.0.
This result is consistent with former observations \citep[e.g.][]{Pierce-Price2000,Molinari2011} indicating that the dust temperatures are uniformly $\sim20$ K throughout the CMZ.
No significant excess in short wavelengths is detected.
The mass of the central clump is estimated to be $5.8\times10^2 \Msol$ from this dust temperature and the 1.1 mm continuum luminosity (0.38 Jy) measured for the entire clump with our SMA data.
We obtain the mean density $n_0 = 5\times10^5\ \pcc$ from the mass and  size of the clump by roughly assuming the volume to be $\frac{4}{3}\pi r^3$.
This is consistent with the density estimate obtained from the excitation analysis of the HCN intensities \citep{Tanaka2014}.

We estimate the viral mass (\Mvt) of the central clump by using the formula $\Mvt/\Msol = 210\,(r/{\mathrm pc})\cdot(\dv/\kmps)^2$ used in \cite{Miyazaki2000} for the CS clumps in the CMZ.
The radius $r$ is estimated as $\frac{1}{2}D\sqrt{\theta_{\mathrm{maj}}\cdot \theta_{\mathrm{min}}} = 0.15\ \pc$, where $D=8.3$ {\bf kpc} is the distance to the source and $\theta_{\mathrm{maj,\ min}} = 10''.2,\ 6''.6$ are the major and minor axis lengths corrected for the instrumental broadening due to the finite beam size, respectively. 
The FWHM velocity width $\dv$ is measured to be $12.4\ \kmps$ from the averaged \NNHp\ spectrum, after correction for broadening due to unresolved hyperfine structures.
These values yield  $\Mvt = 4.8\times10^3\ \Msol$.
The virial parameter $\avir \equiv \Mvt$/$M$ is $8.3$, which is only slightly smaller than the typical value for the CMZ clouds, $\sim 10$, estimated from the CS \JJ{1}{0}\ observations by \cite{Miyazaki2000}.
By considering the rather large uncertainties in dust and gas properties, we conclude that the central clump is not significantly different from the typical CMZ clouds in \avir.

\begin{deluxetable}{lc}
\tablecaption{Properties of the Central Clump and the Methanol Peaks}



\tablecolumns{2}
\small
\tablehead{ \colhead{Parameter} & \colhead{Value} }
\tabletypesize{\scriptsize}
\tablewidth{0pt}
\startdata
\sidehead{Entire Clump}
Dust temperature ($\Td$) & $18.4\pm0.4$ K \phn \\
Radius ($r$)\tablenotemark{1,2} & $0.15\pm0.01$ pc \phn \\
FWHM velocity width (\Dv)\tablenotemark{1,3} & $12.4\pm0.2$ \kmps \phn \\
Dust luminosity mass ($M$) & $5.8\times10^2$ \Msol \\
Virial mass ($\Mvt$) & $4.8\times10^3\ \Msol$ \\
Mean density & $5\times10^5\ \pcc$ \\
\sidehead{Mehtanol Peak-A}
Methanol rotation temperature & $28\pm1$ K \phn\\
Gas density\tablenotemark{4} & $10^{6.5-7.5}\ \pcc$ \\
\sidehead{Mehtanol Peaks-B--H}
Methanol rotation temperature & $28\pm14$ K \phn
\enddata
\tablenotetext{1}{From gaussian fitting of the \NNHp\ image.}
\tablenotetext{2}{Corrected for the beam broadening.}
\tablenotetext{3}{Corrected for the broadening by unresolved hyperfine structures. }
\tablenotetext{4}{By assuming the gas kinetic temperature to be 40--100 K.}


\label{TABLE3}
\end{deluxetable}

\subsection{Methanol Peaks}
\begin{deluxetable}{lrrrrrrrrrr}
\tablecaption{Line Fluxes toward Methanol Peaks in Jy \kmps}



\tablecolumns{9}
\small
\tablehead{ \colhead{} & \multicolumn{4}{c}{\methanol} & \colhead{$\NNHp$} & \colhead{o-\HHCS} & \colhead{\OCS} & \colhead{\CtS}\\
\cline{2-5} \\
\colhead{}&\colhead{\meta}&\colhead{\metb}&\colhead{\metc}&\colhead{\metd}&\colhead{3--2}&\colhead{$8_{1,7}$--$7_{1,6}$}&\colhead{23--22}&\colhead{6--5}}
\tabletypesize{\scriptsize}
\tablewidth{0pt}
\startdata
A   &  $10.6\pm1.4$\phn  &  $60.4\pm1.3$\phn  &  $9.1\pm1.3$\phn  &  $51.0\pm1.3$\phn  &  $125.6\pm1.5$\phn  &  $14.6\pm1.0$\phn  &  $9.8\pm1.6$\phn  &  $9.6\pm1.4$\phn \\ 
B   & \nodata &  $17.5\pm1.1$\phn  & \nodata &  $12.2\pm1.3$\phn  &  $11.4\pm1.6$\phn  & \nodata & \nodata & \nodata \\   
C   & \nodata &  $16.7\pm1.6$\phn  & \nodata &  $15.9\pm1.9$\phn  &  $6.2\pm0.9$\phn  & \nodata & \nodata & \nodata\\   
D   & \nodata &  $22.7\pm2.0$\phn  & \nodata &  $19.7\pm1.9$\phn  &  $11.3\pm1.3$\phn  & \nodata & \nodata & \nodata\\   
E   & \nodata &  $7.8\pm1.6$\phn  & \nodata &  $10.0\pm1.9$\phn  &  $9.4\pm2.0$\phn  & \nodata & \nodata & \nodata\\   
F   & \nodata &  $6.2\pm0.9$\phn  & \nodata &  $10.0\pm1.1$\phn  &  $8.7\pm1.3$\phn  & \nodata & \nodata & \nodata\\   
G   & \nodata &  $10.1\pm1.1$\phn  & \nodata &  $8.8\pm1.2$\phn  &  $9.5\pm1.3$\phn  & \nodata & \nodata & \nodata\\   
H   & \nodata &  $15.8\pm1.7$\phn  & \nodata &  $14.9\pm1.3$\phn  &  $4.1\pm1.2$\phn  & \nodata & \nodata & \nodata\\  
\hline
B--H (stacked) & 
{$1.3\pm0.3$}\phn &
{$13.8\pm0.6$}\phn &
{$2.0\pm0.3$}\phn &
{$12.8\pm0.6$}\phn &
{$8.5\pm0.8$}\phn &
{$3.9\pm0.5$}\phn &
{$2.3\pm0.4$}\phn &
\nodata
\enddata

\label{TABLE4}
\end{deluxetable}

\begin{figure}[ttt]
\epsscale{0.7}
\plotone{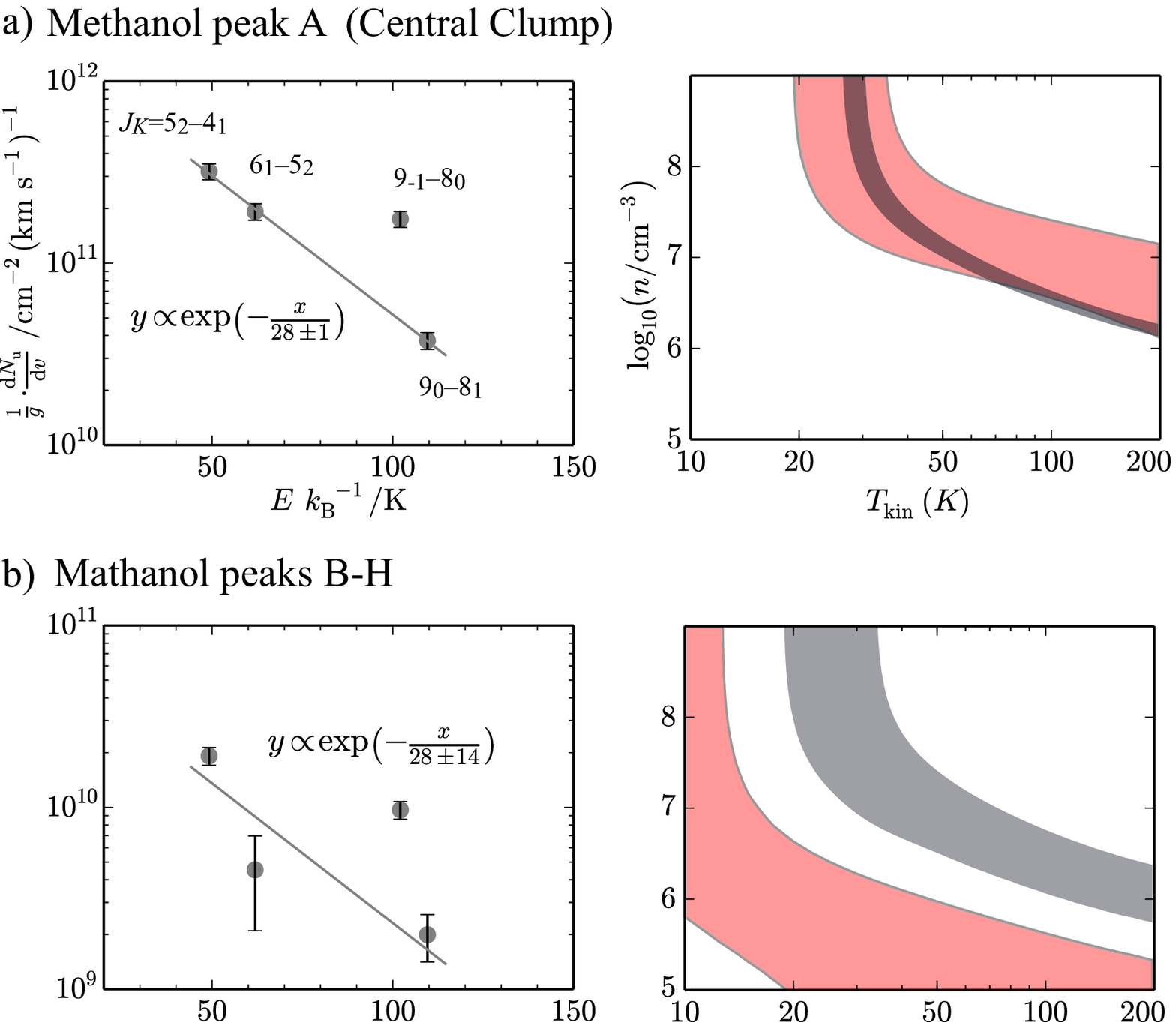}
\caption{(a) Methanol rotation diagram toward peak A, with the best-fit curve of a single-temperature LTE analysis overlaid (left), and the results of non-LTE excitation analysis, in which the parameter ranges calculated from the observed flux density ratios of $J_K=\meta$ and \metc\ to \metb\ are hatched in red and gray, respectively (right).  In the latter analysis, we assume $10\%$ relative error for each flux density in addition to the spectral noise levels.  (b) Same as (a) but for the stacked data for the methanol peaks B--H.}\label{FIG8}
\end{figure}

{\bf
Table \ref{TABLE4} lists the fluxes of the lines detected toward the methanol peaks A--H.
Toward the methanol peaks outside the central clump, i.e., peaks B--H, only  two methanol transitions, \metb\ and \metd, and the \NNHp\ line are detected above the 3-$\sigma$ levels.
We use a stacking analysis to estimate the fluxes of the undetected lines toward peaks B--H; their spectra are co-added after correction for the difference in the central velocities.
All the lines except \CtS\ \JJ{5}{4}\ are detected above the 3-$\sigma$ levels in the stacked spectra, the fluxes of which are listed in the bottom line of Table \ref{TABLE4}. 
}

The left column of Figure \ref{FIG8} shows methanol rotation diagrams constructed from the spectra toward the methanol peak A and the spectra stacked for peaks B--H, in which the upper state column densities per unit velocity (${\rm d}N_{\rm u}/{\rm d}v$) divided by the statistic weights ($g_{\rm u}$) are plotted against their excitation energies for the four methanol transitions.
The ${\rm d}N_{\rm u}/{\rm d}v$ values are calculated from averaged flux densities over the three central velocity channels around the \metb\ peaks by assuming the optically thin condition. 
For the central clump, the flux densities of three out of the four transitions are successfully fitted by a single-temperature local thermodynamical equilibrium (LTE) model with a rotation temperature ($T_{\rm rot}$) of 28 K, but the remaining \metd\ transition has an inconsistently high value possibly because the \metd\ transition is masing.
\cite{Yanagida2014}\ detected {\bf class-I maser emissions} of this methanol line that are associated with molecular outflow sources in the infrared dark cloud G34.43+00.24 MM3.
{\bf Peak A also coincides with a source in the list of 36 GHz class-I methanol maser candidates by \cite{Yusef-Zadeh2013b}.
Therefore, we exclude the \metd\ transition in our data from the excitation analysis, although it is yet to be confirmed whether the transition is actually masing because its spectral profile is similar to those of thermal lines and lacks sharp peaks typical for maser profiles.}
For the stacked data of the methanol peaks B--H, a single temperature LTE fit is less successful, and the uncertainty in the rotational temperature is accordingly larger.  
The best fit $T_{\rm rot}$ value is again 28 K.

We perform a non-LTE excitation analysis to estimate the physical conditions of the methanol peaks by assuming the optically thin condition.
The rate coefficients for the radiative and collisional transitions are taken from the Leiden atomic and molecular database \citep{Schoier2005}.
The right column of Figure \ref{FIG8} shows the ranges of gas kinetic temperature (\Tkin) and molecular hydrogen density (\nHH) that reproduce $\pm 1\ \sigma$ values around the measured flux density ratios among the \meta, \metb, and \metc\ transitions.
In this analysis, we conservatively assume 10\%\ relative errors in the flux densities (i.e. 14\% in the ratios) in addition to the spectral noises. 
In the result for peak A, the \meta/\metb\ and \metc/\metb\ ratios give a consistent parameter range, showing that $\nHH \sim 10^{6.5\mbox{--}7.5}\ \pcc$ with a reasonable assumption for the gas temperature, $40 < \Tkin < 100\ \kelvin$ \citep{Martin2004,Nagai2007}.
For the stacked data for the methanol peaks B--H, the parameter ranges calculated from the two flux density ratios do not overlap.
If we adopt the result for the \metc/\metb\ ratio, which has a higher S/N ratio than the other ratio, we obtain a \nHH\ range similar to that deduced for peak A. 
{\bf The parameters derived in this subsection are listed in Table \ref{TABLE3}.}


\section{DISCUSSION}
\subsection{Kinematics of CO$-0.30-0.07$}
In our previous paper \citep{Tanaka2014}, we presented four hypotheses for the origin of the broad velocity emission of \theObj: cloud--cloud collision, expansion driven by blast wave, molecular outflow, and rotation around an invisible massive object.
We discuss these hypotheses in this subsection.

\subsubsection{Cloud--Cloud Collision}\label{DISCUSSION1}
\begin{figure}[ttt]
\epsscale{0.5}
\plotone{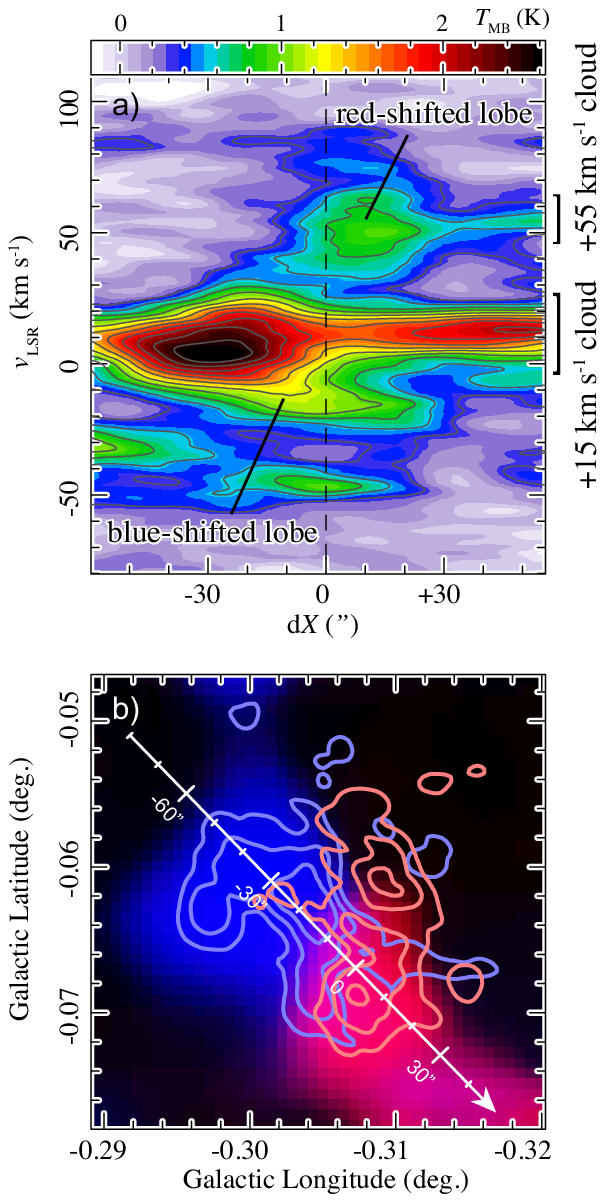}
\caption{(a) Position--\vlsr\ diagram of the ASTE \COt\ \JJ{3}{2}\ data along the path defined in panel b, with contours drawn at every 2 $\sigma$ starting from 3 $\sigma$ ($\sigma = 0.11\ \kelvin$).  The velocity ranges of the +15-\kmps\ and +55-\kmps\ clouds are indicated.  (b) Composite color image constructed from the \COt\ \JJ{3}{2}\ integrated intensities of the +55-\kmps\ and +15-\kmps\ clouds (red and blue color channels, respectively).  Contours of the peak HCN \JJ{3}{2}\ flux densities of the red- and blue-shifted lobes in the SMA data are overlaid and shown in red and blue, respectively. The contours are drawn at 3 $\sigma$\ intervals starting from 7 $\sigma$ ($\sigma = 0.28\ {\mathrm{Jy\,beam^{-1}}}$).}\label{FIG9}
\end{figure}
With the interferometric HCN \JJ{3}{2}\ image taken with the SMA, we found remarkably tight anti-correlation between the two broad velocity lobes comprising \theObj, which is much tighter than that previously found with single-dish ASTE observations;
the characteristic broad velocity emissions are predominantly confined in the central ridges running through the cloud center, and the individual ridges in the two velocity components are tightly anti-correlated with each other on the plane of the sky. 
Thus the entire cloud appears to be sharply divided into the two velocity components by the linear boundary defined by the central ridges.

We can give a natural explanation for this structure by assuming that the cloud is a pair of colliding clumps.
In the numerical model of colliding clouds by \cite{Gilden1984}, planar compressed layers appear behind the shock fronts propagating away from the collision interface and into the individual clouds. 
The central ridges of \theObj, which run along the interface between the two velocity components, are consistent with this model.
The turbulent velocity in the post-shock region is expected to be enhanced up to the shock velocity \citep{Inoue2013b}.
For \theObj, the minimum value of the collision velocity \vcol\ is roughly estimated to be $20\ \kmps$, i.e., half of the  \vlsr\ difference between the two velocity components, although the edge-on geometry of the collision interface may indicate that the tangential collision velocity is faster than the line-of-sight velocity. 
The velocity dispersion of 18\ \kmps\ inside the red-shifted lobe is consistent with this collision velocity within the uncertainty in the tangential velocity.

We can identify a spatially extended part of each of the colliding pair in the single-dish \COt\ \JJ{3}{2}\ map. 
Figure \ref{FIG9}a is the \PV\ diagram of the \COt\ \JJ{3}{2}\ data along the diagonal path across the cloud indicated in Figure \ref{FIG9}b.
The figure shows multiple components in the velocity range from $-50\ \kmps$ to $+90\ \kmps$ in addition to the broad emissions of \theObj.
The blue-shifted lobe of \theObj\ is obviously associated with the +15-\kmps\ cloud, the most intense and most widespread component in the map.
The red-shifted lobe also accompanies a spatially extended, narrow component at $\vlsr\sim+55\ \kmps$.
This +55-\kmps\ cloud is elongated from the south-western direction and is truncated at the position where it intersects the red-shifted lobe.
Figure \ref{FIG9}b shows a composite color image constructed from the single-dish \COt\ data, in which the +15- and +55-\kmps\ clouds are shown in blue and red, respectively.
The figure illustrates that the broad velocity lobes observed in the SMA data (indicated by the contours in the figure) are well correlated with the boundary between the +15- and +55-\kmps\ clouds.
This configuration, i.e., compact broad emissions at the contact points of two clouds with narrower line widths, is typical for a cloud--cloud collision system \citep[e.g.][]{Duarte-Cabral2011}.

A major argument against the cloud--cloud collision scenario is that a part of the shocked region is actually extended well behind the interface between the two clouds in our data; the blue-shifted lobe also has bright emissions in the eastern part of the cloud, where it lacks a red-shifted counterpart (Figure \ref{FIG3}). 
In order to explain this feature, we have to adopt additional, ad hoc assumptions such as that the red-shifted counterpart is somehow invisible or has already dissipated or that there is another source of turbulence in the eastern side of the +15-\kmps\ cloud.

\subsubsection{Expanding Molecular Ring Model}\label{DISCUSSION2}
\begin{figure*}[ttt]
\epsscale{1.}
\plotone{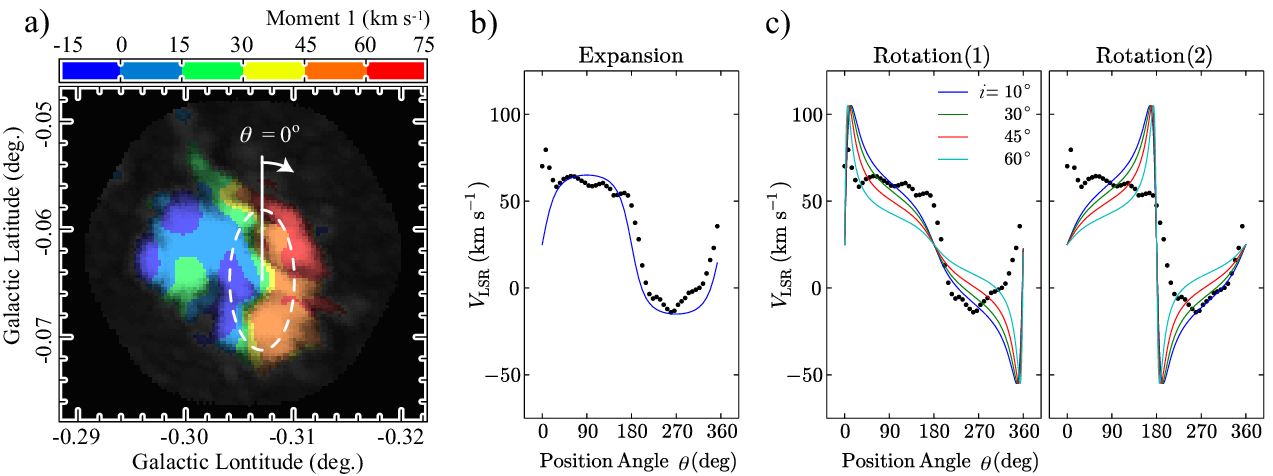}
\caption{a) The first moment map of the SMA HCN \JJ{3}{2}\ data, with the colors and brightness representing the first velocity moment and peak flux density, respectively.  (b) Angular variation of the first moment along the ellipse defined in panel a. The definition of the position angle $\theta$ is also given in panel a.  The overlaid curve is for an expanding-ring model with an expansion velocity $\vexp$ of 45\ \kmps and inclination angle of $62^\circ.5$.  (c) Same as (b) but with model curves for rotating-ring models with various assumptions for the inclination angle $i$.  The maximum line-of-sight velocity is fixed at 80 \kmps.  The attractor loci are assumed to be the northern and southern focal points in models-1 (left) and -2 (right), respectively.}\label{FIG10}
\end{figure*}
An expanding ring driven by a blast wave is also a good model for the cloud kinematics, if we ignore the spatial anti-correlation between the central ridges by regarding it as coincidental.
In the ring model, the two ridges are assumed to be portions of an ellipse elongated in the Galactic north--south direction, as indicated in Figure \ref{FIG10}a.
The figure also shows the first velocity moment map of the SMA HCN \JJ{3}{2}\ data.  
The cloud has an overall velocity gradient across the major axis of the ring, which is interpreted as expansion rather than rotation by adopting the natural assumption that the ring is nearly circular in the three-dimensional space with a large inclination from the plane of the sky.
The other possible kinematics, that is, a model of a highly elliptical rotating ring inclined around the minor axis, is discussed in the next sub-subsection (\ref{DISCUSSION3}).

Figure \ref{FIG10}b\ shows the angular variation of the first moment of the ring defined in Figure \ref{FIG10}a, along with the model curve for an expanding ring with an inclination of $62^\circ.5$ and expansion velocity of $45\ \kmps$. 
Although we have failed to detect clear signs of coherent expansion motion in the \PV\ diagrams (Figures \ref{FIG2} and \ref{FIG6}), the first moment variation in Figure \ref{FIG10} is successfully fitted by the expansion model; both the model and observation show steep gradients at the tangential points ($\theta\sim0^\circ,\ 180^\circ$) and rather flat curves in the other parts.
The failure in the detection of coherent expansion in the three-dimensional \PV\ images could be explained as being the result of disturbance caused by interaction with non-uniformly distributed outer media; such a situation would be expected for an expanding ring driven by explosion in the vicinity of a dense molecular cloud. 

An advantage of the expanding-ring model over the collision model is that it can easily explain the feature unassociated with the central ridges without additional assumptions because an isotropic explosion can create shocked regions also outside the expanding ring. 
A problem with the expansion model is that the region inside the assumed ring is actually not an emission cavity; it shows low-level emissions smoothly bridging the velocity gap between the central ridges, as shown in Figure \ref{FIG5}.   
The central ridges are most smoothly bridged at their midpoints by a horizontal structure, making the ring appear to be shaped like the symbol $\Theta$.
The expanding-ring model cannot provide a simple explanation for the horizontal bar structure.
In addition, the absence of visible energy sources capable of driving the fast expansion motion, the kinetic energy of which is estimated to be $4\times10^{49}\ \ergs$ from the total mass of \theObj\ \citep[$10^3\ \Msol$;][]{Tanaka2014} and the expansion velocity (45 $\kmps$), remains as a major problem; {\bf the cloud is not associated with any sources in the published catalogs for radio \citep{Nord2004,Lazio2008} or X-ray \citep{Muno2006} sources.}
We also note that the first moment variation displayed in Figure \ref{FIG10}b\ is insufficient to distinguish between the collision model and expanding-ring model because the observed pattern is expected for any cases involving two distinct velocity components of similar sizes along the path.

\subsubsection{Other Scenarios}\label{DISCUSSION3}
The other two models in \cite{Tanaka2014}, namely the rotating-ring/disk and the hyper-energetic outflow scenarios, do not fit the observation.
If we assume that the central ridges represent an elliptic Keplarian orbit around a compact massive object, the large eccentricity implies that the attractor is very close to either end of the major axis and that very steep velocity gradients should appear at the pericenter passage as a result.
In Figure \ref{FIG10}c we show two rotation models corresponding to the two possible attractor loci (i.e., the northern and southern focal points) with various inclinations and a fixed maximum line-of-sight velocity of 80 \kmps. 
Neither model reproduces the observed line of sight velocities, especially those near the pericenter passage.
The models with low inclinations, which show milder velocity variation near the pericenter, may seem to fit the observation better, but they require unrealistically large attractor masses: $1\times10^6\ \Msol$ and $1\times10^7\ \Msol$ for inclinations of $30^\circ$ and $10^\circ$, respectively.

The bipolar-outflow model is also unsuccessful; the broad emissions are not linearly aligned  as expected from the outflow model. 

\subsubsection{The Origin of the Broad Emissions}
We have shown that the cloud--cloud collision model gives the most reasonable explanation for the cloud kinematics and the absence of visible energy sources with fewer assumption than the other three models.
We can safely rule out the rotating-ring and  bipolar-outflow models with our observations.
The expanding-ring model cannot be completely excluded, because it reproduces the observed velocity structure of the cloud very well for a limited spatial range. 
Further investigation of the shock structure with a better resolution will help to determine whether the shocked gas represents a collision interface or is shaped as an elongated expanding ring.

\subsection{Implications of the Shock-Triggered Star Formation}
Our SMA observations toward \theObj\ have revealed a remarkable structure quite different from that known from previous observations: a compact, quiescent clump with bright \NNHp\ and continuum emissions co-exists with the highly turbulent gas traced by the HCN and \HCOp\ lines.
We also detected at least 8 point source-like, bright methanol peaks, including one in the central clump.
The densities of the methanol peaks are deduced to be $\sim 10^{6.5\mbox{--}7.5}\ \pcc$ from the excitation analysis with the multi-transition methanol data.
The high density, compactness in size ($\lesssim 0.1\ \pc$ in radius), and bright methanol emissions suggest their similarity to typical star-forming cores.
Hence, we consider that \theObj\ may provide an ideal test case for star formation in the CMZ, where stars generally form under highly turbulent environments relative to the Galactic disk region \citep{Kruijssen2014,Rathborne2014}.

The central clump and all of the methanol peaks are associated with broad HCN emissions, especially those aligned along the boundary between the two velocity components (Figure \ref{FIG4}).
This suggests that they are density enhancements created via compression by the shock formed at the boundary, irrespective of whether the origin of the shock is cloud--cloud collision or blast-wave impact. 
This assumption is corroborated by the widths of the methanol lines; they are  approximately 2--3 times that expected from their sizes and the size--linewidth relationship for the CMZ \citep{Miyazaki2000, Oka2001a}.
The theory of turbulence-regulated star formation \citep{MacLow2004,Krumholz2005,Padoan2011} suggests that the density enhancements in turbulent clouds start collapsing once they become self-gravitating and eventually form stars.
\cite{Inoue2013a}\ show that  high turbulent pressure and enhanced magnetic field strength in cloud--cloud collision systems set a preferable environment for the formation of massive stars.
We may expect such a triggered star formation process in \theObj.

As indicated by the high \avir\ of $\sim 10$, the entire central clump is not self-gravitating but likely confined by external ram pressure.
It is unclear whether the central clump and methanol peaks have any self-gravitating substructures capable of forming stars. 
Here we investigate whether the methanol peak A in the central clump is sufficiently dense to become self-gravitating, according to the modeling in \cite{Federrath2012}.
The critical density \ncrit\ is given by 
\begin{eqnarray}
n_{\rm crit}/n_0 &\sim& A\cdot\alpha_{\rm vir}\cdot\mathcal{M}^2\cdot\left(1+\beta^{-1}\right)^{-1},  \label{EQ1}
\end{eqnarray}
where $n_0$ and $\mathcal{M}$ are the mean density of the cloud and r.m.s. sonic Mach number of the turbulence, respectively, and $A$ is a constant close to unity.
The value $\beta \equiv 2c_{\rm s}^2/c_{\rm A}^2$ is the thermal-to-magnetic pressure ratio, where $c_{\rm s}$ and $c_{\rm A}$ are sound speed and Alfv\'en velocity, respectively.
If the magnetic field is negligible ($\beta\rightarrow\infty$), the critical density is $8\times10^8\ \pcc$ for the central clump when \Tkin = 50 K.
We obtain a lower critical density of $\sim 2\times10^7\ \pcc$ by assuming that $c_{\mathrm{A}} = 3\ \kmps$ from the typical values of the density \citep[$\nHH = 10^{3.5}\ \pcc$; ][]{Nagai2007} and magnetic field strength \citep[$B\sim0.1$ mG;][and references therein]{Ferriere2009} in the CMZ clouds and that the magnetic field is frozen-in to the gas.
The latter value is comparable to the density deduced for peak A, implying that peak A is possibly dense enough to be self-gravitating, although we will need spatially resolved observations toward the methanol peaks in order to make a more conclusive evaluation.
{\bf
Here, we note that the strong tidal field in the CMZ, which is not taken into account in Equation (\ref{EQ1}), does not significantly change the above discussion.
The critical density for self-gravity to overcome the tidal shear is approximately $10^4\ \pcc$ at a Galactocentric radius of 100 pc \citep{Yusef-Zadeh2007}, indicating that the tidal shear is negligible in the density range considered.
}

It may seem that stars are already forming in the methanol peaks in \theObj\ because of the richness of the molecular species detected toward them; bright methanol emission is a major indicator of massive star formation, and \OCS\ and \HHCS\ are also {\bf classified as} hot-core-type molecules \citep{Immer2014a}.
{\bf 
However, we refrain from identifying the methanol peaks as hot cores because the hot-core-like chemistry is not limited to star-forming regions in the CMZ; the hot core chemistry without hot cores \citep{Requena-Torres2006} is rather widespread throughout the region, where the molecules formed on the grain surfaces are efficiently evaporated into the gas phase by cosmic-rays and/or shocks in addition to heating by young massive stars \citep{Requena-Torres2006,Menten2008,Yusef-Zadeh2013b}.}
We estimate the fractional methanol abundance to molecular hydrogen toward peak A to be $2\times10^{-7}$ on the basis of the methanol and continuum flux densities by assuming an ortho/para ratio of unity and the LTE condition with a $\Trot$ of 28 K.
{\bf 
This high abundance is typical for hot cores but is also well within the range of the abundances measured toward non-star-forming clouds in the CMZ, suggesting that the high methanol abundance toward the methanol peaks might simply reflect the typical chemical composition in the CMZ rather than indicate hot-core formation. 
The class-I methanol maser candidates in the \metd\ transition could be further evidence for massive star formation, but it is also possible that they are excited directly by the shock impact rather than by outflow from young stellar objects. 
In fact, the 36 GHz class-I methanol maser candidates identified in \cite{Yusef-Zadeh2013b} show poor correlation with known star-forming regions.}
Internal heating sources are also undetected in the dust SED of the central core (Figure \ref{FIG7}b).

The collision-induced cluster formation scenario is a popular explanation for the formation of super star clusters (SSCs) in the CMZ \citep{Hasegawa1994,Higuchi2014}.
\theObj\ is a new candidate for the site of triggered star formation at its early stage, although it is evident that the cloud, the mass of which is $\sim 10^3\ \Msol$, is only capable of forming a cluster of at most a few $100\ \Msol$.
We note that the situation of \theObj\ is similar to that of G0.253+0.016, or the ``brick'', the infrared dark cloud in the CMZ often referred to as a progenitor of a SSC similar to the Arches and the Quintuplet \citep{Longmore2012a,Higuchi2014,Rathborne2014a}.
\cite{Higuchi2014}\ show sulfur monoxide (SO) cores with large velocity dispersion ($\sigma_v$) aligned along an arc-shaped structure, which they argue is an imprint of past impact by a smaller cloud.
The methanol peaks confined in a thin shocked layer in \theObj\ could be smaller versions of these large-$\sigma_v$ SO cores.

\section{SUMMARY}
By utilizing the SMA, we have performed synthesis mapping toward the HVCC \theObj, a peculiar molecular cloud with an extremely broad velocity width in the CMZ.
We summarize the main results below.
\begin{enumerate}
\item The cloud has two distinctive components: a highly turbulent, spatially extended emission traced by HCN and \HCOp\ and a central clump of 0.15 pc radius with bright \NNHp\ and continuum emissions.  The central clump has a narrow velocity width of $12\ \kmps$ in FWHM, which is in striking contrast to the extremely broad emission of the extended component spanning over a 145 \kmps\ velocity range.   
\item The extended component has a pair of parallel vertical ridges running through the center of the cloud.  These central ridges are tightly anti-correlated with each other in both space and velocity, thereby dividing the entire cloud into two components with line-of-sight velocities differing by 40 \kmps.  
\item The marked broad emissions of the cloud primarily arise from the central ridges, which are again resolved into a number of smaller broad emission features with widths of 40--50\ \kmps\ in the HCN image. These broad HCN emission features are rather randomly distributed in the position--velocity space without apparent order. No pattern indicating systematic motion such as coherent expansion, rotation, or bipolar outflow is detected.
\item The mass of the central clump is estimated to be $5.8\times10^2$ \Msol\ from the continuum luminosity. The virial parameter \avir\ is not significantly different from those of typical CMZ clouds, $\sim10$.
\item At least 8 bright methanol peaks are detected toward the central clump and in the central ridges, accompanying several other hot-core-type molecular lines.  Their methanol rotation temperatures are estimated to be 28 K under the LTE assumption.  Non-LTE analysis for the methanol peak of the central clump shows that the molecular hydrogen density is $10^{6.5\mbox{--}7.5}\ \pcc$ under the assumption that $\Tkin = 40\mbox{--}100\ \kelvin$.
\item \theObj\ is located at a position where two larger clouds at $\vlsr = +15$ and +55 \kmps\ contact each other, and the central ridges with broad emissions are well correlated with the boundary between them.  The tight anti-correlation between the ridges suggest that \theObj\ could be a site of cloud--cloud collision.  Meanwhile, another model in which the central ridges are assumed to be parts of a molecular ring expanding at a velocity of 45\ \kmps\ cannot be excluded, as it explains the angular variation of the line-of-sight velocity well.  The other two models proposed in \cite{Tanaka2014}, namely the bipolar-outflow and the rotating-ring models, do not fit the observations.
\item The central clump and the methanol peaks, which have small radii ($\lesssim 0.1\ \pc$), densities marginally high enough to be self-gravitating, and rich abundances of hot-core-type molecules, are similar to typical star-forming cores in the Galactic disk region.  These objects are predominantly aligned along the central ridges and hence are likely to be features created by shock impact.  We speculate that the cloud is at an early phase of shock-triggered star formation, although conclusive evidence for star formation activities is yet to be obtained.
\end{enumerate}

\acknowledgements
The authors are grateful to an anonymous reviewer for his/her helpful comments.
We also thank the SMA staff for their support during the observations reported here. 
The Submillimeter Array is a joint project between the Smithsonian Astrophysical Observatory and the Academia Sinica Institute of Astronomy and Astrophysics and is funded by the Smithsonian Institution and the Academia Sinica.


%
%

\end{document}